%% file: main.tex
\documentclass[10pt]{article}
\usepackage[utf8]{inputenc}
\usepackage{amsmath,setspace,geometry}
\usepackage{amsfonts}
\usepackage[shortlabels]{enumitem}
\usepackage{graphicx}
\usepackage{bbm}

\usepackage[colorlinks,citecolor=purple,urlcolor=blue,bookmarks=false,hypertexnames=true]{hyperref}
\usepackage[]{natbib} 
\bibpunct[:]{(}{)}{,}{a}{}{,}
\usepackage[english]{babel}
\usepackage{float}
\usepackage{subfig}
\usepackage{booktabs}
\usepackage{pdfpages}
\usepackage{threeparttable}
\usepackage{lscape}

\title{Unified Merger List in the Container Shipping Industry from 1966 to 2022: A Structural Estimation of M\&A Matching}
\author{Suguru Otani\thanks{Market Design Center, University of Tokyo. Email: suguru.otani@e.u-tokyo.ac.jp}\quad  Takuma Matsuda\thanks{Faculty of Commerce, Takushoku University. Email: tmatsuda@ner.takushoku-u.ac.jp}}
\date{
First version: October 15, 2023\\
Current version: \today
}

\begin{document}

\maketitle

\begin{abstract}
We construct a novel unified merger list in the global container shipping industry between 1966 (the beginning of the industry) and 2022. Combining the list with proprietary data, we construct a structural matching model to describe the historical transition of the importance of a firm's age, size, and geographical proximity on merger decisions. 
We find that, as a positive factor, a firm's size is more important than a firm's age by 9.858 times as a merger incentive between 1991 and 2005.
However, between 2006 and 2022, as a negative factor, a firm's size is more important than a firm's age by 2.013 times, that is, a firm's size works as a disincentive.
We also find that the distance between buyer and seller firms works as a disincentive for the whole period, but the importance has dwindled to economic insignificance in recent years. 
In counterfactual simulations, we observe that the prohibition of mergers between firms in the same country would affect the merger configuration of not only the firms involved in prohibited mergers but also those involved in permitted mergers.
Finally, we present interview-based evidence of the consistency between our merger lists, estimations, and counterfactual simulations with the industry experts' historical experiences.

\end{abstract} 

\vspace{0.1in}
\vspace{0in}

\section{Introduction}

Container shipping plays a pivotal role in global trade, revolutionizing the world. According to data from IHS Markit and Descartes Datamyne, it constituted 45.4\% of amount-based imports to the U.S., 21.3\% of amount-based exports from the United States, and 10.1\% of quantity-based world trade in 2021.
Moreover, the container shipping industry presents an intriguing opportunity to investigate industry dynamics, including entry, exit, and investment \citep{otani2023industry}, as well as the history of mergers since its global shipping operations inception in 1966.
Despite its significance, there exists a notable absence of a consolidated dataset for container shipping mergers, particularly from 1966-1990, hindering quantitative research in this area. 
This study addresses this gap by providing a unified list of all realized mergers in the container shipping industry from 1966 to 2022.
We highlight not only our data contribution in itself but also a novel structural estimation analysis to study an endogenous merger incentive and its historical transition in the container shipping industry, which is anecdotally described (\cite{broeze2002globalisation}, \cite{levinson2016box}).

Using our new merger list, we depict the merger waves in the global container shipping industry from 1966 to 2022 and compare them with the price and quantity transitions constructed by \cite{matsuda2022unified}.
The first merger wave emerged following the enactment of the Shipping Act of 1984. 
Subsequently, significant merger waves occurred after 2005, aligning with the exponential growth in quantities under competitive prices. With targeted periods of the data sources, this observation leads us to categorize the industry's history into three distinct ``regimes" corresponding to the data:--  1966-1990, 1991-2005, and 2006-2022.
Additionally, by merging our merger list with proprietary ship-level data, we delineate the merger patterns, revealing a tendency for mergers to involve relatively younger and smaller firms in more distant countries in recent years.
For example, Taiwan's Cheng Lie Navigation, which was acquired by France's CMA-CGM in 2006, was a relatively small container shipping company founded in 1971. 
Safmarine, which was acquired by Denmark's Maersk in 1999, was a South African shipping company. 
The data patterns are consistent with the view that, in addition to mergers and acquisitions, the market share of independent operators in the liner shipping industry has declined in recent years \citep{Merk2022MEL}. 

The merger patterns provide crucial insights into merger waves in the global container shipping industry.
Multiple factors can account for this pattern. 
For instance, recent acquisitions may prioritize the firm's size to expand market tonnage shares, unlike the period from 1966 to 1990. 
Alternatively, companies may emphasize geographical proximity to establish dominance at the local level.
To untangle these explanations and gain more precise insights into various channels, we employ a structural model that quantifies the relative significance of a firm's age, size, and geographical proximity using a novel approach — the matching maximum score estimator \citep{fox2018qe} as in the merger analysis of \cite{akkus2015ms}.

Our estimation results indicate that the assortativeness of both size and geographical proximity contributes to merger incentives or disincentives. 
Overall, firm size influences merger incentives or disincentives in different periods, while geographical proximity consistently contributes to disincentives across all periods.
First, the assortativeness of a firm's size shifts from positive (1991-2005) to negative (2006-2022). 
During 1991-2005, the importance of a firm's size supersedes the importance of its age by a factor of 9.858, serving as a merger incentive.
Conversely, between 2006 and 2022, as a negative factor, a firm's size is more important than a firm's age by 2.013 times, that is, a firm's size works as a disincentive.
In addition, we observe that geographical distance acts as a merger disincentive throughout the entire period, albeit its relative importance compared to the firm's age has dwindled to economic insignificance in recent years. 
This finding suggests diminished merger incentives for establishing dominance at the local country level.\footnote{Our findings, from a practical perspective, indicate that the relationship between shipping companies and their respective home country governments weakens. This change is largely a result of globalization, driven by the expansion of transportation networks following the advent of containerization. This global expansion has altered traditional practices, reducing the direct influence of national governments on shipping companies. For example, in the post-World War II era, as pointed out by \cite{Heidbrink2012}, we observe the increasing tendency of shipping companies to separate the flag state of their vessels from the countries of actual ownership. This shift emerged as the use of flags of convenience became a widespread standard in the international maritime business, allowing shipping companies to operate more globally and reducing the direct control and influence of their home countries. Furthermore, since the 1980s, there has been significant progress in regulatory enhancements concerning shipping alliances \citep{matsuda2022unified}. This regulatory change is part of a broader trend of diminishing state control over shipping companies. Our dataset includes examples of several state-owned shipping companies being privatized or sold, including South Korea's KSC, Australia's ANL, France's CGM, and Singapore's NOL. In Japan, although the government orchestrated the merger of shipping companies in 1964 \citep{otani2021estimating}, it had no involvement in the formation of the Ocean Network Express \citep{hatano2023}, further illustrating this trend of reduced government involvement.}

Finally, we conduct a counterfactual simulation based on the estimated parameters to examine the consequences of prohibiting mergers between firms within the same country.
This type of merger restriction is highly contentious in competition policies of the global market, particularly in global container shipping.
For instance, on June 21, 2017, South Africa's Competition Commission issued a statement forbidding the consolidation of container businesses through three shipping lines — Nippon Yusen Kaisha(NYK), Mitsui O.S.K. Lines (MOL), and Kawasaki Kisen Kaisha (KLINE). 
The commission expressed concerns about market consolidation by domestic companies and cartel issues related to these firms in the car carrier business.
Although the country's competition court eventually approved the integration on January 17, 2018, its potential impact on the planned launch of the integrated container company Ocean Network Express remains noteworthy.
In our counterfactual simulations, we discover that prohibiting mergers between firms in the same country affects not only the merger outcomes of the involved firms but also those engaged in permitted mergers. 
This foretells the ripple effect of local competition policies through equilibrium matchings, influencing both local markets and the global market configuration. 
We also provide interview-based evidence of the consistency between our merger lists, estimations, and counterfactual simulations with the industry experts' historical experiences.

\subsection{Related literature}

This study contributes to three strands of the literature: empirical transferable utility (TU) matching, endogenous merger analysis, and recent industrial policy and antitrust studies in the shipping industry.

First, it contributes to the literature on empirical TU matching. 
See the recent methodological development in \cite{agarwal2021market}.
The most related econometric model is \cite{fox2010qe,fox2018qe}, whose model has been applied to other empirical topics such as banking mergers \citep{akkus2015ms,chen2013ijio}, faculty room allocations \citep{baccara2012aer}, executive and firm matchings \citep{pan2017determinants}, and buyer and seller relationships in the broadcast television industry \citep{stahl2016aer}. 
These papers have applied the matching maximum score estimator proposed by \cite{fox2010qe,fox2018qe} to two-sided many-to-many and one-to-one matching in a TU matching environment. 
We apply this approach to merger waves in the global container shipping industry from its inception by dividing its history into three regimes based on institutional knowledge and data periods.
To the best of our knowledge, our paper is the first to illustrate historical the transitions of assortativeness of observed variables using long panel data.

Second, it contributes to the literature on endogenous merger analyses. 
Endogenous merger analysis in the industrial organization literature can be divided into dynamic and static matching models. 
In terms of dynamic matching models, they mainly follow \cite{gowrisankaran1999dynamic}.\footnote{\cite{stahl2011dynamic} is the first to estimate a merger activity model using a dynamic, strategic framework. \cite{jeziorski2014effects} estimates the sequential merger process to analyze ownership consolidation in the United States radio industry after the enactment of the Telecommunications Act of 1996. \cite{igami2019mergers} apply a stochastic sequential bargaining model to the merger processes of the hard disk industry. As the most recent paper, \cite{hollenbeck2020horizontal} enriches the Gowrisankaran-type dynamic endogenous merger model. With different dynamic approaches, \cite{nishida2015better} compare post-merger and pre-merger beliefs and equilibrium behaviors in a Markov perfect equilibrium in the Japanese retail chain industry. \cite{perez2015building} incorporates mergers as bidding games by incumbents and investigates the effect of the Reagan-Bush administration's merger policy on the reallocation of assets in the United States cement industry.} 
Conversely, in terms of static matching models, \cite{uetake2019entry} develop an empirical two-sided non-transferred utility matching model with externalities using moment inequalities and investigate the effect of entry deregulation on the ``with whom"-decisions of bank mergers by the Riegle-Neal Act. 
\cite{akkus2015ms} address the same Act with a different approach. 
They add transfer data and construct a one-to-one matching model with transfer utility and find that merger value increased from cost efficiencies in overlapping markets, relaxing regulations, and the network effects exhibited by acquirer-target matching. 
Our paper follows \cite{akkus2015ms} and focuses on endogenous mergers in a single, static, large matching market for each regime, quantifying the relative importance of tonnage capacity and geographical proximity, which are the main economic forces driving firms to pursue mergers to gain cost efficiency in the shipping industry \citep{notteboom2004container}.
In addition, we compare the historical transitions of the relative importance of the variables to derive the potentially different merger incentives behind merger waves in the industry.

To the best of our knowledge, no prior studies have comprehensively examined all mergers in the container shipping industry. Our merger dataset stands as the initial benchmark, making a significant contribution to maritime transportation literature. The only relevant paper in the industrial organization literature is \cite{akkus2015ms}. Our findings for the period 1966-1990 align with their observation that large acquirer banks tend to match with larger target banks in the U.S. banking industry. However, our findings for the period 2006-2022 reveal new features distinct from those reported by \cite{akkus2015ms}.

Third, this study contributes to the literature on recent industrial policy and antitrust in the shipping industry. In the industrial organization literature, \cite{jeon2022learning} studies the relationship between learning and investment in the container shipping industry between 2006 and 2014 and simulates social welfare in counterfactual merger scenarios in which a merger occurred between the top two firms that jointly account for over 35\% of total capacity in the industry.
In the maritime shipping literature, various researchers use the Herfindahl-Hirschman Index (HHI) and its modification, although empirical studies using simple regressions of HHI are criticized \citep{bresnahan1989empirical}.\footnote{As a theoretical merger analysis, \cite{nocke2022concentration} demonstrate that in a general Cournot model, only the naively-computed change in the HHI due to a merger (twice the product of the per-merger market shares of the merging firms), but not the level of the HHI, is valuable in screening mergers for welfare assessment. Out of the merger analysis, \cite{spiegel2021herfindahl} shows that the HHI reflects the ratio of producer surplus to consumer surplus in Cournot markets under some theoretical conditions. The empirical maritime literature follows the direction of \cite{spiegel2021herfindahl} focusing on the relationship between market concentration, prices, and consumer surplus without taking care of the endogeneity problem of prices, HHI, etc, for example, \cite{Hirata2017}. This has been criticized in industrial organization literature \citep{bresnahan1989empirical} since the 1980s, so the literature has developed a conduct parameter estimation approach to recover markups and the extent of collusion. See the original model \citep{bresnahan1982oligopoly} and theoretical and numerical properties \citep{matsumura2024conduct,matsumura2023test,matsumura2023revisiting}. Policymakers and practitioners often use the HHI to measure the market concentration of the container shipping market. For example, the Federal Maritime Commission (FMC), the administrative agency that oversees shipping in the United States, cited the competitive nature of the HHI as one of a reason why the Transpacific route is extremely competitive \citep{FMC2022}.} For example,
\cite{Sys2009TransPOL} collects firm-year-level vessel volume data for the period 1999-2009 and calculates the HHI and Gini coefficients to compare the degree of market concentration. 
The author shows that while the degree of concentration has increased in years when mergers and acquisitions have taken place, the industry is still fragmented and competitive due to small shares of firms.
\cite{Merk2022MEL} use a modified Herfindahl–Hirschman Index (MHHI) to show that the industry concentration is higher when consortia are considered.
Although these papers treat specific hypothetical mergers and consortia as exogenously determined scenarios and focus on the post-merger market outcomes such as some welfare and concentration measures based on non-cooperative game theoretical models, our paper endogenizes mergers based on cooperative game-theoretical models, i.e., matching models, and focus on merger incentives. 
Our approach enables us to simulate hypothetical endogenous mergers based on inferred merger incentives instead of being unable to assess welfare and concentration measures as in the above studies.

The remainder of this paper is organized as follows. 
Section \ref{sec:data_and_institutional_background} summarizes the data and institutional background of mergers in the container shipping industry.
Section \ref{sec:empirical_analysis} constructs a structural matching model to quantify the assortativeness of the observed characteristics for each regime and compare the levels across regimes.
Section \ref{sec:results} presents our estimation results.
Section \ref{sec:counterfactuals} shows counterfactual simulation results.
Section \ref{sec:practical_implications} summarizes the practical implications, discussions, and directions for future research.
Finally, conclusions are presented in Section \ref{sec:conclusion}.

\section{Data and Industry Background}\label{sec:data_and_institutional_background}
We provide the details of the data sources in Section \ref{sec:data_source}. 
We provide industry background in
Section \ref{sec:industry_background} and summary statistics for the variables in Section \ref{sec:descriptive_statistics}.

\subsection{Data source}\label{sec:data_source}
We compile data by merging three distinct sources.
The initial source is the Containerization International Yearbook (CIY), which offers ship-level information from 1966 to 1990. 
The second source is the IHS Markit data (IHS), which provides ship-level information from 1991 to 2005. 
The third source, the Handbook of Ocean Commerce (HB), provides ship-level data from 2006 to 2022.
We classify these periods in the respective data sources as ``regimes," resulting in three regimes: 1966-1990, 1991-2005, and 2006-2022.
By consolidating ship-level data, we construct firm-year-level variables, including country names and owned operating container capacity, measured in Twenty-foot Equivalent Units (TEU), which is available as a common capacity variable across our three data sources.
Finally, we manually create a merger list containing the buyer and seller names along with the merger years. 
This list is subsequently integrated with the firm-year-level variables using institutional information. 
Note that MDS Transmodal data is a potential alternative for a fee, but it offers a maximum of two years of panel raw data, which includes ship-year level variables used in our analysis.
Therefore, we believe that our data construction is the best and most feasible.

We have some remarks because we find some inconsistencies such as a one-year lag and missing ship-level variables between these data sources and institutional facts.
First, we fix the inconsistencies following the observations in the newer regime data. 
Second, we treat the firms not operating in the merged year as firms with constant capacity levels from the last active year in the merged year. 
Third, we treat mergers of container shipping seller firms by non-container-shipping firms outside the industry as exits from the container shipping sector, as these mergers lack information on buyer firms.
Fourth, we treat consolidation-type mergers as mergers in which buyer firms have the lower bound of age and size variables at the initial merger timing.
The final data on mergers are summarized in this section and used in empirical analysis in Section \ref{sec:empirical_analysis}.

\begin{figure}[!ht]
\begin{center}
  \includegraphics[width = 0.7\textwidth]
  {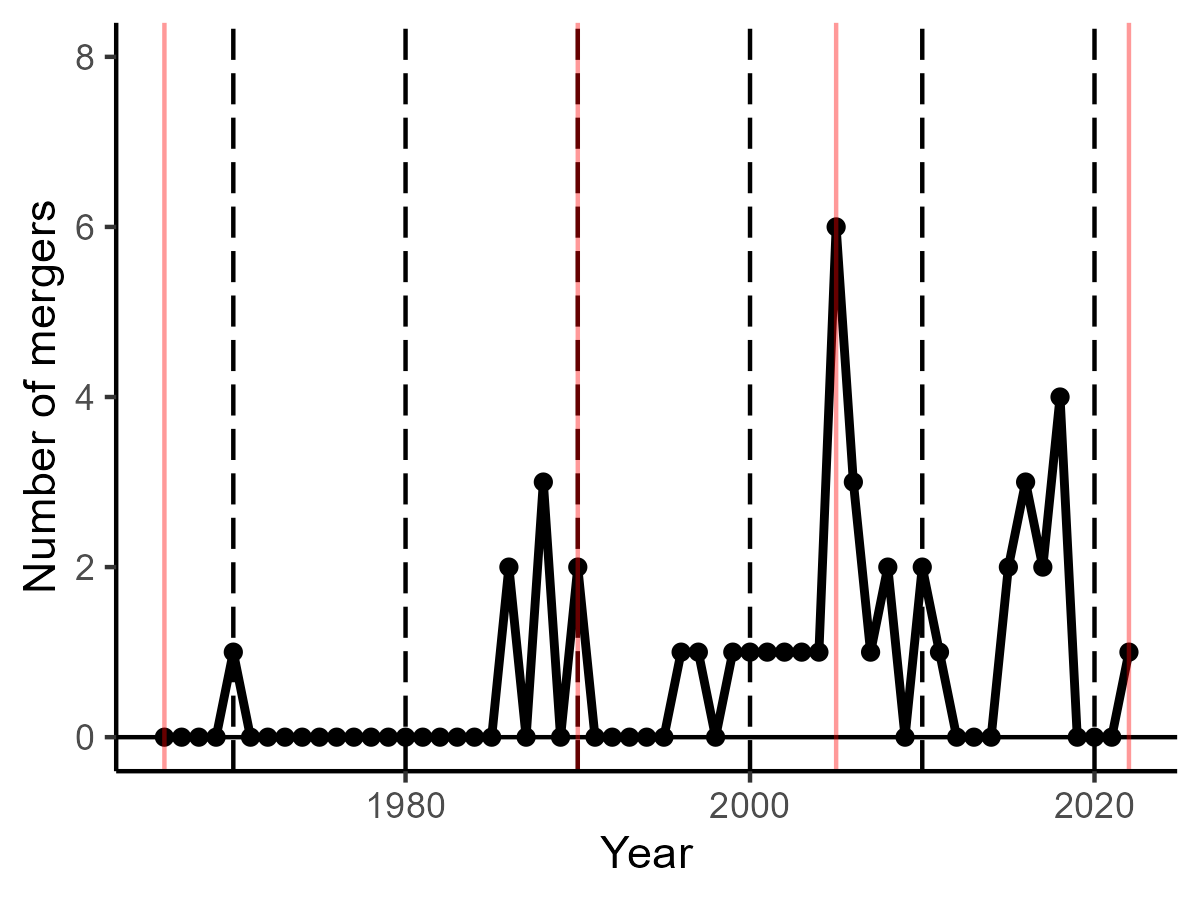}
  \caption{The number of mergers between 1966 and 2022}
  \label{fg:number_of_mergers}
  \end{center}
\footnotesize
   Note: Red lines divide the regimes based on the CIY, IHS, and HB. We treat consolidation-type mergers as mergers in which buyer firms have the lower bound of age and size variables at the initial merger timing.
\end{figure}

\begin{figure}[!ht]
\begin{center}
  \subfloat[Price]{\includegraphics[width = 0.7\textwidth]
  {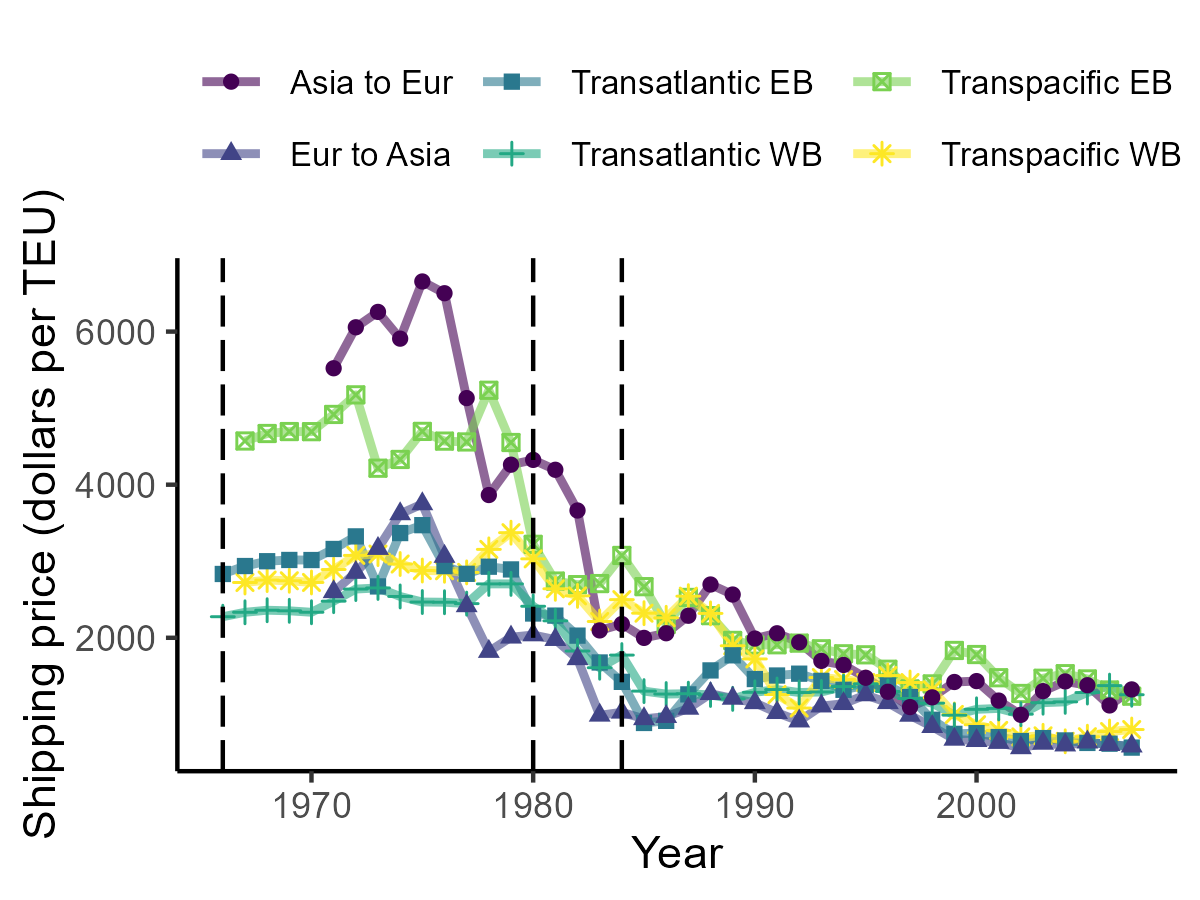}}\\
  \subfloat[Quantity]{\includegraphics[width = 0.7\textwidth]
  {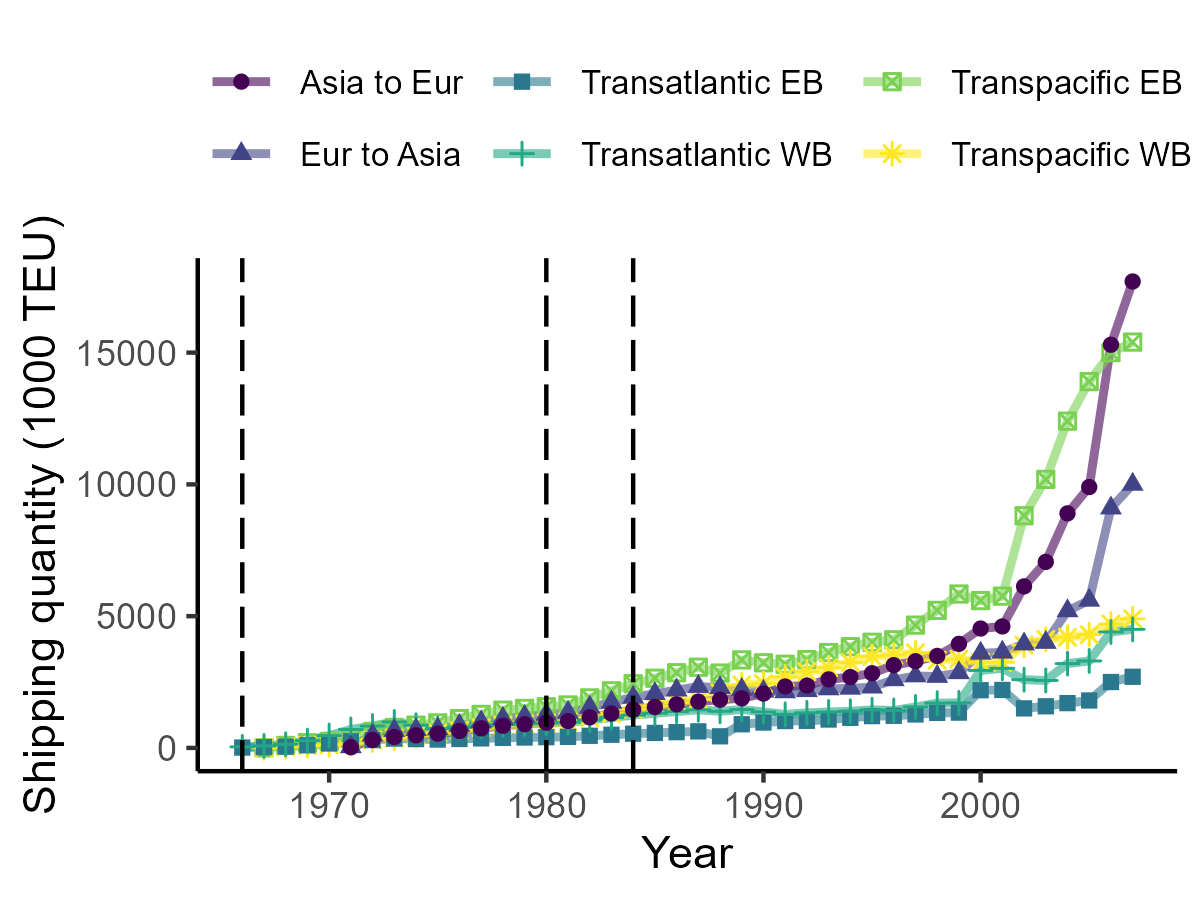}}
  \caption{Trends in route-year-level shipping prices and quantities.}
  \label{fg:container_freight_rate_and_shipping_quantity_each_route}
  \end{center}
\footnotesize
  Note: Prices are adjusted to the CPI in the U.S. in 1995. Dotted vertical lines show the timing of cartel breakdown events of the Container Crisis. See details in \cite{matsuda2022unified}.
\end{figure}

Figure \ref{fg:number_of_mergers} illustrates the number of mergers between 1966 and 2022 based on a merger list.
For comparison, Figure \ref{fg:container_freight_rate_and_shipping_quantity_each_route} illustrates the trends in route-year-level shipping prices and quantities between 1966 and 2009.
Comparing these figures provides graphical intuition. 
First, merger waves emerged after the enactment of the Shipping Act of 1984, signifying a shift from collusive behavior.
Second, subsequent merger waves align with the exponential growth in quantities under competitive prices post-2005.
As a result, we categorize the industry's history into three distinct ``regimes": 1966-1990, 1991-2005, and 2006-2022, in accordance with the corresponding data.

\subsection{Industry Background}\label{sec:industry_background}
We chronologically describe the industry background between 1966 and 2022  by focusing on firms' mergers.\footnote{The M\&A deals presented in this paper include shipping companies operating across various sectors beyond container shipping, such as COSCO, CHINA SHIPPING, Japanese companies, and Maersk in the energy sector. However, since our research focuses on mergers in the container transport sector, this inclusion is justifiable as we rely solely on container ship data. Integrating data from other sectors is infeasible due to the limitations of our dataset. Additionally, we use operated fleet capacity as our primary metric, as it aligns closely with sales volumes and provides a more suitable indicator of company size compared to owned capacity.} 
As mentioned earlier, we divide the entire period into three distinct regimes—1966-1990, 1991-2005, and 2006-2022 — aligned with institutional background and data sources explained later.
In the subsequent merger lists, we deliberately retain the original firm names for each regime's data, despite potential inconsistencies across these datasets.

Note that mergers and acquisitions differ in whether the target company survives, and are essentially different things in terms of corporate decision-making. However, as in the analysis in this paper, as long as the main focus of the analysis is on the scale of service provision and the transport network itself, it can be seen that there is no significant difference between merger and acquisition in the container transport business.

\begin{table}[!htbp]
  \begin{center}
      \caption{Merger list: CIY (1966-1990)}
      \label{tb:merger_list_CIY} 
      \input{figuretable/merger_list_CIY}
  \end{center}\footnotesize
  Note: We treat consolidation-type mergers as mergers in which buyer firms have the lower bound of age and size variables at the initial merger timing.
\end{table} 

\begin{table}[!htbp]
  \begin{center}
      \caption{Merger list: IHS (1991-2005)}
      \label{tb:merger_list_IHS} 
      \input{figuretable/merger_list_IHS}
  \end{center}\footnotesize
  Note: We omit the merger of Royal Nedlloyd and the Peninsular and Oriental containers (P\&O Containers) in 1997, NYK and Showa Kaiun in 1998, Maersk and Sea-Land in 1999, and CMA and CGM in 1999 because historical vessel data of merged firm are missing and we could not identify these mergers from our data. We treat consolidation-type mergers as mergers in which buyer firms have the lower bound of age and size variables at the initial merger timing.
\end{table} 

\begin{table}[!htbp]
  \begin{center}
      \caption{Merger list: HB (2006-2022)}
      \label{tb:merger_list_HB} 
      \subfloat[HB (2006-2022)]{\input{figuretable/merger_list_HB}}\\
      \subfloat[HB (2006-2022): Inconsistent merger cases]{\input{figuretable/merger_list_HB_inconsistent}}
  \end{center}\footnotesize
  Note: In Panel (a), we list the seller of ID21 as APL, which was acquired by NOL in 1997, because the merged entity continued to list APL as the registered operator name in the HB data up until 2017. Additionally, we have identified the buyer of ID13 as Maersk, rather than Hamburg Sud, because Maersk purchased Hamburg Sud in 2018.
  Panel (b) provides merger cases that have inconsistencies with the institutional background because ships operated by merged firms can keep the past operator name after merger years.  In addition, we omit COSCO's merger of OOCL in 2018 because the merged OOCL's vessels keep OOCL as the registered operator name in the HB data until 2022; therefore, we could not identify the renewal timing of the registered operator names. We treat consolidation-type mergers as mergers in which buyer firms have the lower bound of age and size variables at the initial merger timing.
\end{table} 

\paragraph{1966-1990} 

Table \ref{tb:merger_list_CIY} summarizes all mergers based on CIY between 1966 and 1990.\footnote{In 1964, the Japanese ocean shipping industry experienced consolidation induced by the government, and 95 firms were merged into six large groups. \cite{otani2021estimating} investigates the event using a structural matching model.}
This period involves a collusive and competitive environment with shipping conferences that are explicit and cartels globally allowed. 
The period is studied in \cite{matsuda2022unified} and \cite{otani2023industry}.
The period is divided into collusive (1966-1983) and competitive (1984-1990) periods according to the U.S. Shipping Act of 1984.\footnote{The relevant studies are \cite{wilson1991some}, \cite{pirrong1992application}, and \cite{clyde1998market}. \cite{wilson1991some} provided evidence of regime change by the Shipping Act of 1984 using data on quarterly freight rates and shipping quantities of five selected commodities only on the Transpacific route. \cite{pirrong1992application} tested the model prediction of the core theory surveyed in \cite{sjostrom2013competition} using data from two specific trade routes between 1983 and 1985. \cite{clyde1998market} studied the relationship between market power and the market share of shipping conferences after the Act. }
In the collusive period between 1966 and 1983, one merger occurred in 1970 (Moore-McCormack Lines, Inc. merged by United States Lines). 
During the competitive period between 1984 and 1990, two mergers occurred in 1986, three occurred in 1988, and two occurred in 1990. 

Until the 1970s, containerized operations were primarily implemented in developed countries like North America, Western Europe, and Japan \citep{GUERRERO2014151}, while many other countries did not widely adopt containerization.
In addition, container freight rates were high until the 1970s owing to shipping conferences, as shown in Figure \ref{fg:container_freight_rate_and_shipping_quantity_each_route}.
Therefore, aside from the Moore-McCormack merger, there were no significant mergers in the industry during the 1970s because colluding firms were able to generate sufficient profits without the need for mergers and acquisitions. In the newspaper interview, executives from Japanese shipping companies indicated that liner shipping, including container shipping, was the most profitable segment of their companies during that period and played a vital role in supporting their operations \citep{sato2006}.

The 1980s marked a significant expansion of containerization within the industry, primarily occurring in North America, Western Europe, Japan, and their trading partners, including the Caribbean, the Mediterranean, and Southeast Asian countries \citep{GUERRERO2014151}. During this period, these regions were integrated into the global trade network with the beginning of offshoring and the emergence of new transshipment hubs such as Singapore. However, there was a significant decline in container freight rates due to the Sea-Land's withdrawal from shipping conferences in 1980 and the enactment of the Shipping Act of 1984 in the United States \citep{matsuda2022unified}. 
The declining freight rates and the depreciation of the dollar resulting from the Plaza Accord of 1985 significantly impacted the profitability of Japanese and European shipping companies \citep{Matsuda2018}, leading to a series of mergers and acquisitions .
In Japan, Yamashita Shin Nihon Kisen Kaisha (Y-S Line) and Japan Line merged to form the Nippon Liner System (NLS) in the container shipping sector. In 1988, Hanjin Shipping merged with the Korea Shipping Company (KSC), which was originally state-owned. 

\paragraph{1991-2008}

Table \ref{tb:merger_list_IHS} summarizes all the mergers based on the IHS between 1991 and 2005.
This period includes the enactment of the Ocean Shipping Reform Act (OSRA) of 1998 which divides the period into pre-OSRA and post-OSRA periods. 
\cite{fusillo2006some,fusillo2013stability} and \cite{reitzes2002rolling} conducted detailed studies on this period.
We find two mergers before 1998 and twelve mergers after the enactment of the OSRA of 1998.

Container shipping was established as one of the global standards for carrying cargo in the 1990s. 
New ports were developed in East Asian countries including China. Several ports grew as new transshipment hubs, such as Salalah and Colon, to accommodate growth in emerging economies, such as Vietnam, India, and Brazil \citep{GUERRERO2014151}. Despite Panamax size being considered the maximum size for container vessels prior to this period, shipping companies began to employ vessels that exceeded the Panamax limit.

Global alliances in the container shipping were established in the 1990s \citep{Hirata2017}. 
When a shipping company chooses to merge with another company as part of its liner network expansion strategy, it can gain increased market power by owning or chartering a greater number of vessels. However, this also exposes the company to significant risks associated with fluctuations in container freight rates or cargo volumes. If the shipping companies choose to form or join an alliance, it can provide customers with more comprehensive networks without the need for mergers. Nevertheless, it does not have the power to set prices, as members of the alliance make pricing decisions independently.

Shipping alliances involved global-scale cooperation, primarily on major routes such as the transpacific and the Far East to Europe. 
This was in contrast to the cooperation seen in earlier shipping conferences, which were limited to specific routes.
In 1994, MOL formed The Global Alliance (TGA) with APL, Nedlloyd and OOCL.
The following year, NYK, Hapag-Lloyd, NOL and P\&O Containers formed the Grand Alliance (GA).
The CKY Alliance was formed in 1996 by KLINE, COSCO and Yang Ming Shipping. 
Hanjin Shipping entered the CKY alliance in 2001 to form the CKYH alliance.
The Royal Nedlloyd Lines merging with P\&O became P\&O Nedlloyd and decided to join GA in 1997.
After NOL's acquisition of APL, NOL and P\&O Nedlloyd formed The New World Alliance (TNWA) with MOL and Hyundai Merchant Marine in 1998.
Firms that chose not to join alliances sought to expand through mergers. For example, CMA merged with CGM, and Maersk merged with Sea-Land in 1999.

In the first half of the 2000s, merger incentives among shipping companies declined because many firms achieved substantial profits in the transpacific and Far East-Europe markets without resorting to mergers. This was attributed to factors such as increased exports of electrical appliances, furniture, and household goods, driven by China's rapid economic growth, a stable housing market in the U.S., and strong economic growth in Europe. 
Container cargo movements, especially on routes from China to the U.S. and Europe, experienced substantial increase.
As some exceptions, mergers and acquisitions were mainly European companies seeking to increase their scale, including Maersk's merger with P\&O Nedlloyd, Hapag-Lloyd's merger with CP Ships in 2005, and CMA-CGM's merger with Delmas in 2006.

\paragraph{2009-2022}
Table \ref{tb:merger_list_HB} summarizes all mergers based on the HB between 2006 and 2022. 
We find inconsistent merger cases between the HB data and institutional background; therefore, we split the merger cases into two categories.
Panel (a) summarizes the merger cases from the HB data, consistent with institutional history.
In panel (b), we summarize three merger cases from the HB data that are inconsistent with the institutional history.
The inconsistency comes from the fact that merged firms can keep the past operator name in the operation. 
We treat the three merger cases as occurring in the data, based on the data records.
Summarizing the two panels, we find five mergers before 2009 and 16 mergers after the enactment of the OSRA of 1998.

A remarkable feature of this period was the oversupply of shipping services.
After the U.S. housing market collapsed in 2007 with the revelation of the subprime mortgage crisis among financial institutions, transport volume growth slowed. 
On the other hand, the number of container vessels increased. 
Additionally, shipping companies expanded the size of their vessels to lower the operational cost per container.
This encouraged the expansion of service supply and deteriorated the supply-demand balance.
For example, \cite{matsuda2022} show that, compared with 1986, the volume of global container cargo transport was 723\% in 2007 and 1047\% in 2016. However, the total capacity of container vessels increased from 944\% in 2007 to 1784\% in 2016.
The oversupply of shipping services resulted in a significant decrease in freight rates in the late 2010s.\footnote{Falling freight rates also led to alliance restructuring and mergers. In 2012, GA and TNWA began joint operations on Asia-Europe routes, forming the G6 alliance. 
The three European companies, Maersk, MSC, and CMA-CGM, announced the formation of a new alliance called the P3 Network in 2013.However, this was terminated due to lack of approval from the Chinese Ministry of Commerce. This was due to the fact that it would have had too large market share on the Far East-Europe route. 
Following the failure of the P3 alliance, Maersk and MSC immediately signed a 10-year vessel-sharing agreement to form the 2M alliance. The remaining CMA-CGM also announced the formation of Ocean Three (O3) with China Shipping Container Line (CSCL) and Arab-owned United Arab Shipping Company (UASC), and began partnering on major routes in 2015. The CKYH alliance was also joined by Evergreen in April 2014, replacing the CKYHE alliance.} As a result, container shipping firms chose to merge to survive the tough shipping market.

Mergers have made significant progress since 2014. 
That year, Hapag-Lloyd's merger with CSAV, Hamburg Sud's merger with CCNI, and CMA-CGM's merger with the German shipping line ODPR were announced.
In 2015, CMA-CGM merged NOL and Chinese government decided to merge the container shipping divisions of the COSCO Group and China Shipping Group.
In 2016, Hapag-Lloyd's merger with UASC and Maersk's merger with Hamburg Sud were announced, and in Japan, NYK, MOL, and KLINE announced the merger of their liner shipping divisions.
In August 2016, the first bankruptcy of a major shipping company (Hanjin Shipping) occurred since the formation of alliances.
Restructuring in the 2010s has been significant in scale, with eight container shipping companies bankrupt or merged between 2015 and 2018.

\subsection{Descriptive statistics}\label{sec:descriptive_statistics}

Table \ref{tb:summary_statistics} presents summary statistics for firm-year-level variables of buyer and seller firms in all realized merger cases from 1966 to 2022. 
These variables include the firm's age in the global container shipping industry and size measured by TEU in the merger year. 
We normalize these values from 1e-6 (minimum age or size) to 1 (maximum age or size) within each regime for inter-regime comparison.
First, we observe a decline in the mean of normalized firm ages, from 0.84 in the 1966-1990 period to 0.52 in the 2006-2022 period. 
This suggests that recent mergers involve relatively younger firms, that is, firms with less experience in the container shipping industry.

\begin{figure}[!ht]
\begin{center}
  \includegraphics[width = 0.45\textwidth]
  {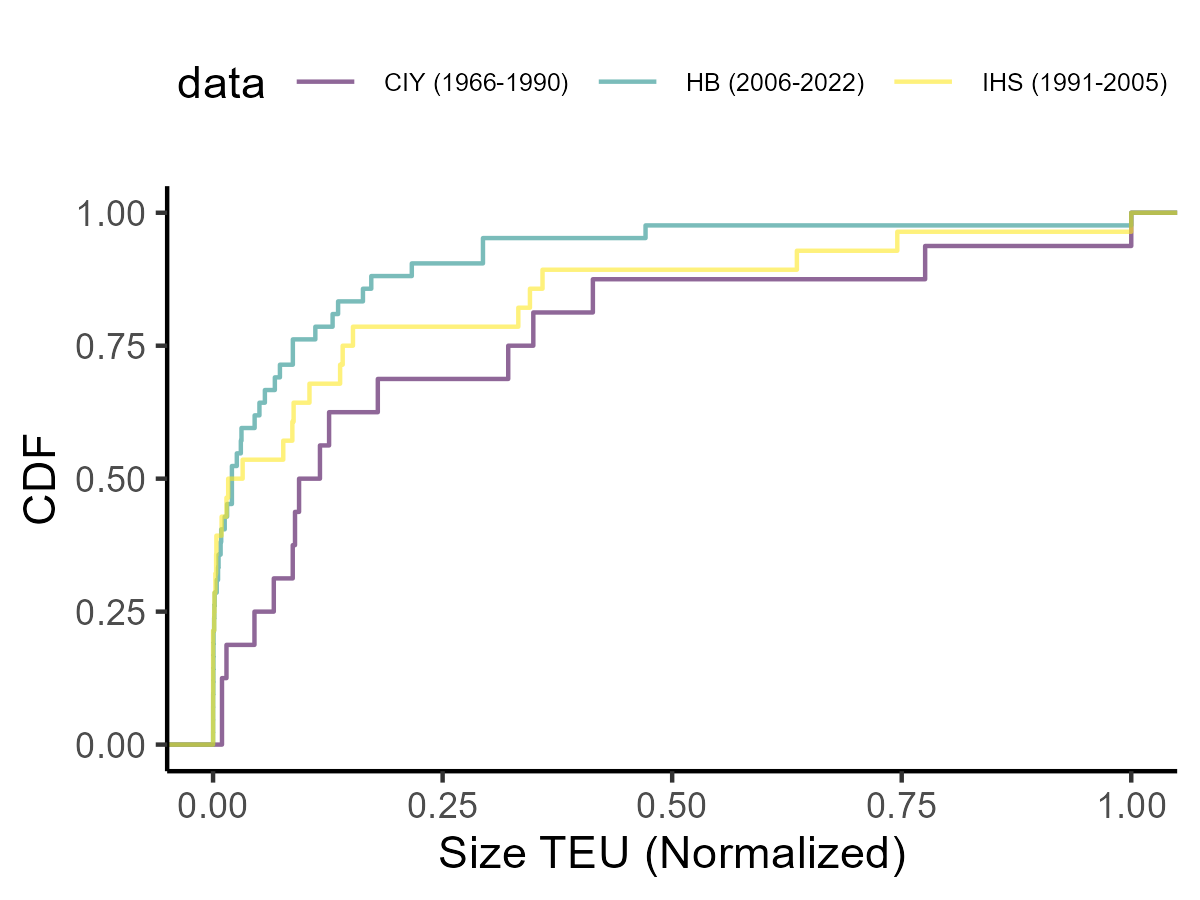}
  \includegraphics[width = 0.45\textwidth]
  {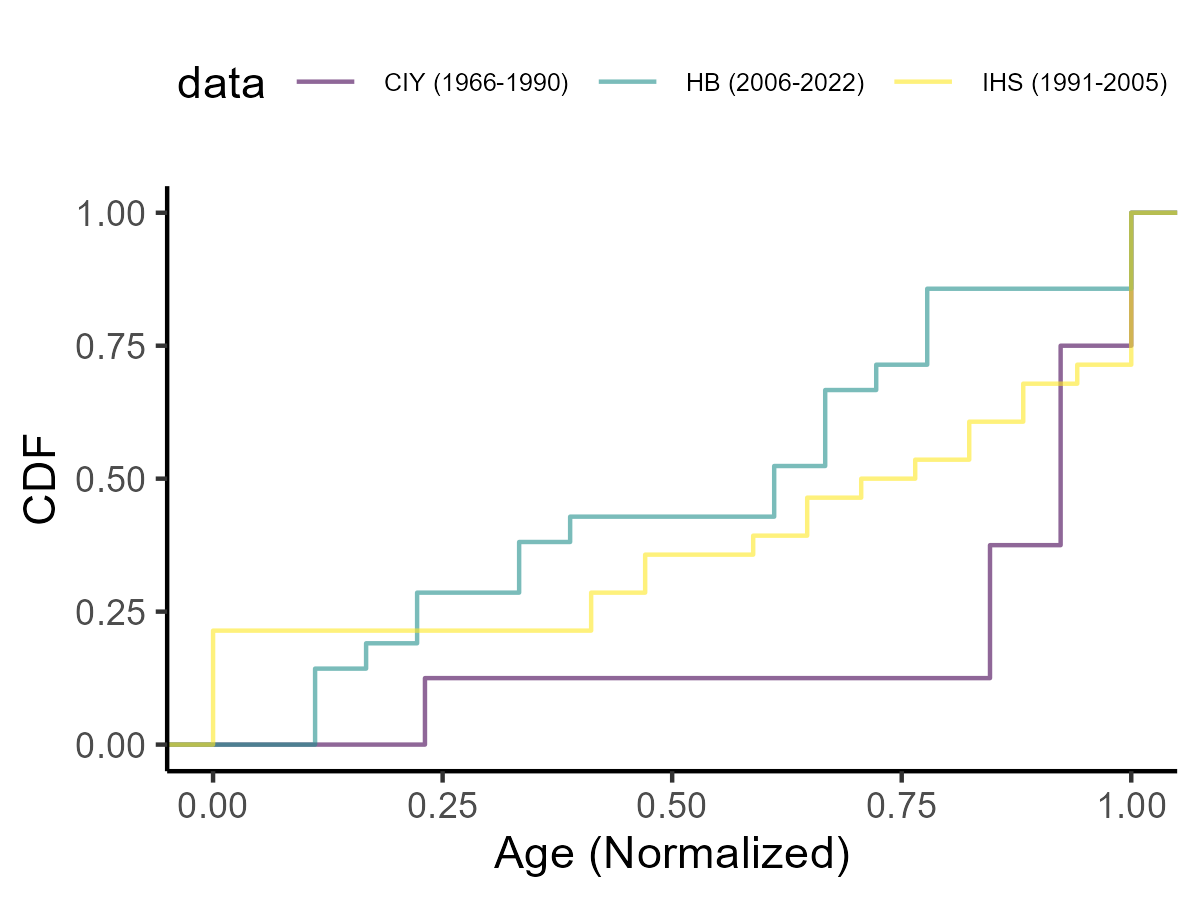}
  \caption{Distributions of firm-year-level variables for each regime}
  \label{fg:size_cdf}
  \end{center}
\footnotesize
\end{figure}

Second, the mean of normalized firm sizes decreases from 0.23 in the 1966-1990 period to 0.07 in the 2006-2022 period.
This indicates that recent mergers involve relatively smaller firms, despite the exponential growth in firm size, particularly from 2006 to 2022.
Notably, we treat new entrant firms purchasing incumbent firms as having an age and size of zero, reflecting the increased number of entries through mergers.
Figure \ref{fg:size_cdf} illustrates the cumulative distributions of normalized firm size and age for each regime, reinforcing the observation that recent mergers tend to feature younger and smaller firms.

\begin{table}[!htbp]
  \begin{center}
      \caption{Summary statistics of firm-year-level variables}
      \label{tb:summary_statistics} 
      \subfloat[CIY (1966-1990)]{\input{figuretable/summary_statistics_of_firms_CIY}}\\
      \subfloat[IHS (1991-2005)]{\input{figuretable/summary_statistics_of_firms_IHS}}\\
      \subfloat[HB (2006-2022)]{\input{figuretable/summary_statistics_of_firms_HB}}
      
  \end{center}\footnotesize
\end{table} 

\begin{figure}[!ht]
\begin{center}
  \includegraphics[width = 0.7\textwidth]
  {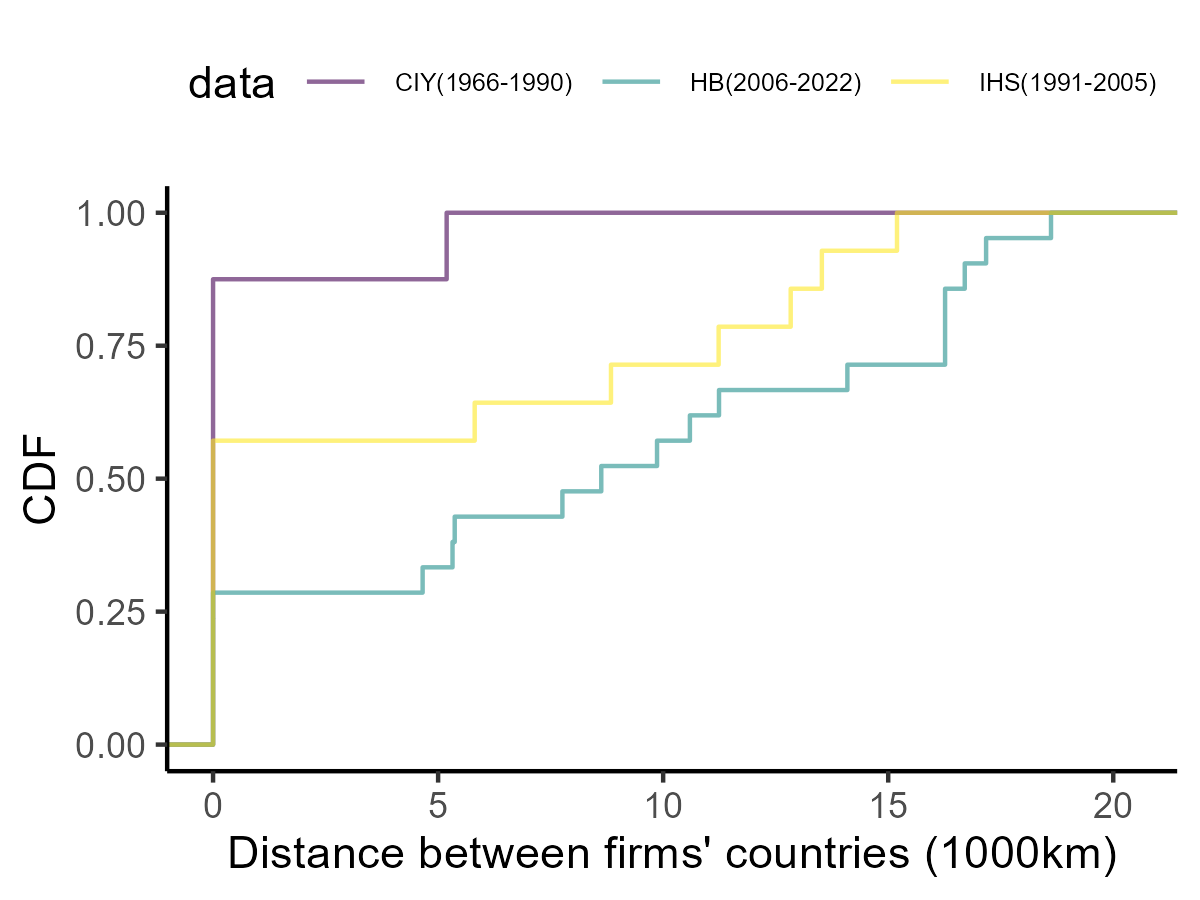}
  \caption{Distributions of match-level distances of seller and buyer firms for each regime}
  \label{fg:distance_cdf}
  \end{center}
\footnotesize
\end{figure}

Figure \ref{fg:distance_cdf} depicts distributions of realized match-level distances calculated as the Euclidean distance based on the latitude and longitude of the capital cities of flag countries of matched firms. If the flag countries are the same, the distance is zero.
We observe that nearly all mergers between 1966 and 1990 involve buyer and seller firms within the same country. 
Although mergers within the same country account for over 30\% of the entire period, the number of mergers between firms separated by distance has increased recently. 
This shift suggests that the reasons for mergers have transitioned from domestic to global, driven by the growth and expansion of international container transport.

These visualizations provide initial insights into the involvement of relatively younger and smaller firms in more distant countries in recent mergers. 
However, comprehensively assessing the relative importance of each variable with limited data requires a sophisticated structural model. 
In Section \ref{sec:empirical_analysis}, we introduce a structural matching model to quantify the significance of each variable and analyze the transitions across regimes.

\section{Empirical analysis}\label{sec:empirical_analysis}

Our objective is to quantify the assortativeness of observed characteristics for each regime and compare the levels across regimes to illustrate the merger trend of the global container shipping industry. 
We employ a matching maximum score estimator developed by \cite{fox2018qe}, which is a well-known method for measuring matching assortativeness based solely on matching patterns.

We model mergers in each regime as a two-sided one-to-one transferable matching game in a single market. Let $\mathcal{N}_b$ and $\mathcal{N}_s$ be the sets of potential finite buyers and sellers respectively. Let $b=1,\cdots,|\mathcal{N}_b|$ be buyer firms and let $s=1,\cdots,|\mathcal{N}_s|$ be seller firms where $|\cdot|$ is cardinality.\footnote{As shown in Table \ref{tb:summary_statistics} and one-to-one matching feature, $|\mathcal{N}_b|=|\mathcal{N}_s|=8$ for the regime (1966-1990), $14$ for the regime
(1991-2005), and $21$ for the regime (2006-2022).} Let $\mathcal{N}_{b}^{m}$ denote the set of ex-post matched buyers and $\mathcal{N}_{b}^{u}$ denote that of ex-post unmatched buyers such that $\mathcal{N}_b= \mathcal{N}_{b}^{m}\cup\mathcal{N}_{b}^{u}$ and $\mathcal{N}_{b}^{m}\cap\mathcal{N}_{b}^{u}=\emptyset$. For the seller side, define 
$\mathcal{N}_{s}^{m}$ and $\mathcal{N}_{s}^{u}$ as the set of ex-post matched and unmatched sellers such that $\mathcal{N}_s= \mathcal{N}_{s}^{m}\cup\mathcal{N}_{s}^{u}$ and $\mathcal{N}_{s}^{m}\cap\mathcal{N}_{s}^{u}=\emptyset$. 
Let $\mathcal{M}^m$ be the sets of all ex-post matched pairs $(b,s)\in\mathcal{N}_{b}^{m}\times \mathcal{N}_{s}^{m}$. Let $\mathcal{M}$ denote the set of all ex-post matched pairs $(b,s)\in\mathcal{M}^{m}$ and unmatched pairs $(\tilde{b},\emptyset)$ and $(\emptyset,\tilde{s})$ for all $\tilde{b}\in \mathcal{N}_b^u$ and $\tilde{s}\in \mathcal{N}_s^u$ where $\emptyset$ is a null agent generating unmatched payoff. 

Each firm can match at most one agent on the other side; thus,  $|\mathcal{N}_b^{m}|=|\mathcal{N}_s^{m}|$. The matching joint production function is defined as $f(b,s)=V_b(b,s)+V_s(b,s)$ where $V_b:\mathcal{M}\rightarrow \mathbb{R}$ and $V_s:\mathcal{M}\rightarrow \mathbb{R}$. The net matching values for buyer $b$ and seller $s$ are defined as $V_b(b,s)=f(b,s)-p_{b,s}$ and $V_s(b,s)+p_{b,s}$, where $p_{b,s}\in \mathbb{R}_{+}$ is the equilibrium merger price paid to seller firm $s$ by buyer firm $b$ and $p_{b\emptyset}=p_{\emptyset s}=0$. For scale normalization, we assume $V_b(b,\emptyset)=0$ and $V_s(\emptyset,s)=0$ for all $b\in \mathcal{N}_b$ and $s\in \mathcal{N}_s$. Each buyer maximizes $V_b(b,s)$ across seller firms, whereas each seller maximizes $V_s(b,s)$ across buyer firms. 

The stability conditions for buyer firm $b \in \mathcal{N}_b$ and seller firm $s \in \mathcal{N}_s$ are as follows:
\begin{align}
    V_b(b,s) &\ge V_b(b,s') \quad \forall s' \in \mathcal{N}_s \cup \emptyset,s'\neq s,\label{eq:stability_ineq}\\
    V_s(b,s) &\ge V_s(b',s) \quad \forall b' \in \mathcal{N}_b\cup \emptyset,b'\neq b.\nonumber
\end{align}

Based on Inequalities \eqref{eq:stability_ineq} and equilibrium price conditions $p_{b',s}\le p_{b,s}$ and $p_{b,s'}\le p_{b',s'}$ in \cite{akkus2015ms}, we construct the inequalities for matches $(b,s)\in \mathcal{M}$ and $(b',s')\in \mathcal{M}, (b',s')\neq(b,s)$ as follows:
\begin{align}
    f(b,s)-f(b,s')&\ge p_{b,s}-p_{b,s'}\ge p_{b,s}-p_{b',s'},\label{eq:pairwise_stable_ineq}\\
    f(b',s')-f(b',s)&\ge p_{b',s'}-p_{b',s}\ge p_{b',s'}-p_{b,s},\nonumber\\
    V_s(b,s)-V_s(b',s)&\ge 0,\nonumber\\
    V_{s'}(b',s')-V_s(b,s')&\ge 0,\nonumber
\end{align}
where $p_{b',s}$ and $p_{b,s'}$ are the unrealized equilibrium merger prices that cannot be observed in the data. The last two inequalities cannot be derived from the data because the researchers cannot observe how the total matching value $f(b,s)$ is shared between buyer $b$ and seller $s$.

Each buyer firm can only acquire one seller firm, which implies that the buyer firm’s choice among a set of seller firms is discrete. 
As a simple semiparametric technique for estimating the discrete choice, we employ maximum score estimation \cite{manski1975maximum,manski1985semiparametric}.
\cite{fox2018qe} proposes a maximum score
estimator using the above inequalities when we observe the transfer data or not under mild conditions. 
The maximum score estimator is consistent if the model satisfies a rank order property as in \cite{manski1975maximum,manski1985semiparametric}, i.e., the probability of observing matched pairs is larger than the probability of observing swapped matched pairs. 
The rank order property is equivalent to pairwise stability which is a milder property rather than stability, so the rank order property holds under the above stability conditions. See identification details in \cite{fox2010qe} and Monte Carlo simulation results in \cite{fox2018qe}, \cite{akkus2015ms}, and \cite{otani2021matching_cost}.

We specify $f(b,s)$ as a parametric form $f(b,s|X,\beta)$ where $X$ is a vector of observed characteristics of all buyers and sellers and $\beta$ is a vector of parameters. 
In the absence of transfer data, given the observed characteristics $X$, one can estimate $\beta$ by maximizing the following objective function:
\begin{align}
    Q(\beta)=\sum_{(b,s)\in \mathcal{M}} \sum_{(b',s')\in \mathcal{M},(b',s')\neq (b,s)} \mathbbm{1}[f(b,s|X,\beta)+ f(b',s'|X,\beta)\ge f(b,s'|X,\beta)+f(b',s|X,\beta)]\label{eq:score_function}
\end{align}
where $\mathbbm{1}[\cdot]$ is an indicator function. 
The inequality is constructed by adding the first two inequalities and canceling out transfers $p_{b,s}$ and $p_{b',s'}$ in Inequalities \eqref{eq:pairwise_stable_ineq}.
The objective function \eqref{eq:score_function} counts the number of correctly predicted pairwise stable matching under each candidate parameter $\beta$.

In our empirical application, as the observed characteristics, $X$, we use the standardized firm's age, size measured by the total tonnage, and match-level distance calculated from the locations of the flag countries at merger timing, that is, all observed variables are standardized to $[1e-6,1]$. 
Concretely, we specify the joint production function $f(b,s)$ as
\begin{align}
    f(b,s|X,\beta)= \beta_1 \text{Age}_{b}\text{Age}_{s} + \beta_2 \text{Size}_{b}\text{Size}_{s} + \beta_3 \text{Distance}_{bs} + \varepsilon_{bs},\label{eq:joint_production}
\end{align}
where $\varepsilon_{bs}$ is assumed to be i.i.d. errors drawn from the zero median distribution as in \cite{fox2018qe}. 
Note that any parameters of firm-specific characteristics cannot be identified with maximum score estimation based solely on without-transfers information. With transfer data, $p_{b,s}$ and $p_{b',s'}$, such as the payments regarding mergers from buyer firms to seller firms, the identification is possible and the precision of the estimator improves as in \cite{akkus2015ms}.

\section{Results}\label{sec:results}

Table \ref{tb:maximum_score_estimate} reports the estimation results of the matching maximum score estimator. 
As we can use only realized merger cases, we could not construct a 95 \% confidence interval with sufficient subsampled data via bootstrap.
Instead, the numbers in brackets indicate the lower and upper bounds of a set of maximizers of the objective function. 
If the lower and upper bounds are the same, then the parameters are point-identified. 
Otherwise, the parameters are partially identified.
The percent of correct matches used as a measure of statistical fit is more than 90\%, so the estimated model predicts the actual mergers well.\footnote{Using the estimated parameters, we compute all possible inequalities in \eqref{eq:score_function} and evaluate whether each inequality is the same as the corresponding actual inequality in the data.}

First, the estimated coefficient of the firm's size shows an interesting transition. 
The sign is changed from ambiguous between 1966 and 1990, positive between 1991 and 2005, to negative between 2006 and 2022. 
In particular, between 1991 and 2005, as a positive factor, a firm's size is more important than a firm's age by 9.858 times in merger decisions, that is, a firm's size works as a merger incentive.
Conversely, between 2006 and 2022, as a negative factor, a firm's size is more important than a firm's age by 2.013 times, that is, a firm's size works as a merger disincentive.
These results are consistent with the institutional fact that consolidation-type mergers in which buyer firms have the lower bound of age and size variables at the initial merger timing have been common rather than absorption-type mergers in recent years.

The primary motivation for container shipping companies to expand is the pursuit of economies of scale \citep{ITF2019}. The motivation to pursue economies of scale is likely to be greater for larger companies. This is because while the market is growing, the fixed costs required to maintain the size of the fleet and global network are growing even faster \citep{RMT2022}. This motivation was behind the fact that the size of the company acted as an incentive for mergers between 1991 and 2005.

On the other hand, since 2006, the size of the companies has become an obstacle to mergers, and this is thought to be related to the fact that in recent years there has been a trend toward consolidation among container shipping companies, and it has become easier for competition authorities and governments to raise objections when large companies merge. Several reports and articles have noted this relationship. For example, \cite{RMT2022} has noted that it is becoming more difficult for companies with large operations to merge, stating that ``Container ship sizes seem to have reached a maximum and further mergers and acquisitions are constrained by regulatory limitations." When Maersk merged with Hamburg Sud, the merged company was required to withdraw from the consortium on several routes \citep{EU2017}, and when COSCO and CHINA SHIPPING merged, the U.S. government reportedly required them to sell their container terminals \citep{WSJ2019}.

Second, the estimated coefficient of the distance of seller and buyer firms shows a negative sign across all regimes but the level decreases between 2006 and 2022. 
This means that mergers of firms in distant countries are likely to occur, however, the importance level relative to the firm's age has decreased to economically zero recently.
These results are consistent with data patterns shown in Section \ref{sec:descriptive_statistics} and the institutional facts that shipping companies do not hesitate to merge with companies in distant regions to expand their container shipping networks.

The reduced significance of distance as a barrier in the shipping industry is likely due to mergers emphasizing the importance of maintaining a global network. \cite{notteboom2012container} identified one of the key motives for mergers and acquisitions in liner shipping as the opportunity to gain immediate access to markets and distribution networks. Similarly, \cite{yeo2013geography} observed that in the container shipping industry, larger companies are more likely to acquire firms located on a different continent, highlighting the strategic importance of international expansion. Furthermore, \cite{brooks2006mergers}, in their analysis of shipping mergers and acquisitions from 1996 to 2000, reported that 40\% of these transactions were cross-border, underscoring the global nature of such activities.

\begin{table}[!htbp]
  \begin{center}
      \caption{Matching maximum score estimation}
      \label{tb:maximum_score_estimate} 
      \input{figuretable/maximum_score_estimate}
  \end{center}\footnotesize
  Note: The objective function was numerically maximized using the differential evolution (DE) algorithm in \texttt{BlackBoxOptim.jl} package. For the DE algorithm, we require setting the domain of parameters and the number of population seeds so that we fix the former to $[-10, 10]$. For estimation, 100 runs of 1000 seeds were performed for all specifications. The numbers in parentheses are the lower and upper bounds of the set of maximizers of the maximum rank estimator. Parameters that can take on only a finite number of values (here 1) converge at an arbitrarily fast rate, then they are superconsistent. The unit of measure of all variables is normalized to $[1e-6,1]$. 
\end{table} 

\section{Counterfactual}\label{sec:counterfactuals}

In the global container shipping industry, mergers between firms in the same country involve several concerns from competition policies in multiple countries.
For example, on June 21, 2017, South Africa's Competition Commission issued a statement stating that it "forbade" the integration of the container business by the three shipping lines of NYK, MOL, and KLINE. 
The commission cited concerns about market consolidation by domestic companies and cartel issues involving these companies in the car carrier business.
The country's competition court finally approved the integration on January 17, 2018; however, this could impact the planned launch of the integrated container company, Ocean Network Express. 
In another example, COSCO and China Shipping's alliance decision in 2015 could face scrutiny from regulators as the market share in some east-west trades is likely to breach the 30\% mark if the new entity joins the group. Above 30\%, alliances must ensure their agreements comply with EU rules outlawing anti-competitive behavior.

In our counterfactual simulation, given estimated parameters, we simulate the matching outcome under the hypothetical scenario that mergers between firms in the same country are prohibited. 
Mechanically, first, this scenario uses the merger cases of firms in different countries and imposes that the joint production function \eqref{eq:joint_production} is changed to
\begin{align*}
    f(b,s|X,\beta)= \begin{cases}
        \beta_1 \text{Age}_{b}\text{Age}_{s} + \beta_2 \text{Size}_{b}\text{Size}_{s} + \beta_3 \text{Distance}_{bs} + \varepsilon_{bs}, \quad \text{if }\text{Distance}_{bs}\neq 1e-6,\\
        -\infty, \quad \text{otherwise}.
    \end{cases}
\end{align*}
Second, given counterfactual $f(b,s|X,\beta)$, we compute an equilibrium matching allocation. 
The equilibrium one-to-one matching allocation $\{m(b,s)\}_{b\in\mathcal{N}_b,s\in\mathcal{N}_s}$ is calculated according to the following linear programming problem proposed by \cite{shapley1971assignment}:
\begin{align*}
    \max_{\{m(b,s)\}_{b\in\mathcal{N}_b,s\in\mathcal{N}_s}} &f(b,s|X,\beta)\cdot m(b,s),\\
    \text{s.t. } 0&\le \sum_{b\in\mathcal{N}_b}m(b,s)\le 1\quad  \forall s \in \mathcal{N}_s,\\
    0&\le \sum_{s\in\mathcal{N}_s}m(b,s)\le 1\quad \forall b \in \mathcal{N}_b,\\
    0&\le m(b,s) \quad \forall b \in \mathcal{N}_b,\forall s \in \mathcal{N}_s,
\end{align*}
where the dual of this linear programming problem also gives equilibrium prices. 
In the equilibrium matching allocation, $m(b,s) = 1$ if firms $b$ and $s$ are matched and $m(b,s) = 0$ otherwise.
In the simulation, we fix the parameters to the upper bounds of estimated parameters in Table \ref{tb:maximum_score_estimate}, and we draw 100 i.i.d. draws of $\varepsilon_{bs}$ from the standard normal distribution $N(0,1)$, then solve the above linear programming problem.
We confirm that using the lower bounds of parameters gives the same results.
We report the lower and upper bounds of percentages of the number of total matchings and the same matching configurations of the simulated 100 matching outcomes relative to the data.

\begin{table}[!htbp]
  \begin{center}
      \caption{Counterfactual simulations under the prohibition of mergers of firms in the same country}
      \label{tb:number_of_mergers_counterfactual} 
      \input{figuretable/number_of_mergers_counterfactual_upper_bound}
  \end{center}\footnotesize
  Note: We simulate the matching outcome from the upper bounds of the estimated parameters and 100 i.i.d. draws of $\varepsilon_{bs}$ from the standard normal distribution $N(0,1)$. The bracket gives the lower and upper bounds of percentages of the number of total matchings and the same matching configurations of the simulated 100 matching outcomes relative to the data.
\end{table} 

Table \ref{tb:number_of_mergers_counterfactual} presents the counterfactual simulation results under the prohibition of mergers between firms in the same country. 
We compare hypothetical merger cases with actual cases and focus on the two regimes between 1991 and 2005 and between 2006 and 2022, as only one merger of firms in different countries happened between 1966 and 1990.
First, the counterfactual model predicts the number of mergers between firms in different countries perfectly in both regimes. 
Second, as an interesting finding, only 9.5 to 47.6\% of the counterfactual merger pairs are the same as actual pairs for the regime between 2006 and 2022.
This implies that the prohibition would restrict the choice set of buyer firms involved in the prohibited mergers to the set of firms in different countries and change the matching outcome through maximization of the modified objective function \eqref{eq:score_function}, in particular, between 2006 and 2022.
Therefore, we find that the prohibition of mergers between firms in the same country affects the merger configuration of all firms in the industry.

\paragraph{Robustness}
We mention the robustness of our findings. First, regarding estimation, the matching maximum score estimation is a semiparametric approach so robust to the unobserved errors rather than parametric estimation such as regression and likelihood approaches. The distributional assumption is only the zero median error which is a mild condition in econometrics literature shown in \cite{manski1985semiparametric}. In other words, the matching maximum score estimation is robust to unobserved factors such as implicit CEO relationships, preferences of CEO, and so on under the limited data. 
Second, for counterfactual simulations, we check several specifications of the variance of errors and find that our results of counterfactual simulations are unchanged.

\section{Interviews}\label{sec:interviews}
In this section, we present interview-based evidence of the consistency between our merger lists, estimations, and counterfactual simulations with the industry experts' historical experiences. We interviewed Hiroyuki Sato, former Vice-Chairman of Mitsui O.S.K. Lines (MOL) and former Manager of the company's North American Division, who responded to our email inquiry on November 2nd. We also interviewed Yasuhiro Fujita, former Japan Representative of American President Lines (APL), with over thirty years of experience in the container shipping industry, who responded to our email inquiry on November 3rd. Prof. Jong Kil Han, who previously worked at Hyundai Merchant Marine, responded to our email inquiry on October 26th. We inquired about their thoughts and memories related to the figures, tables, and our empirical findings.

\paragraph{Validity of our merger list}

All interviewees mentioned that the lists adequately encompass many significant merger events but have certain limitations.
Sato, Fujita, and Han highlighted the omission of numerous smaller shipping companies in the tables, which, in reality, succumbed to fierce competition and filed for bankruptcy, becoming mere wreckage in the Pacific and Atlantic oceans. 
Sato further emphasized that several mergers in our list are categorized as acquisitions, despite our model not necessitating a distinction between the roles of selling and buying firms. 
Additional specific discrepancies are outlined in Appendix \ref{sec:details_of_data_construction}.

\paragraph{Consistency of our data pattern and estimation results}
In our data, we observe that companies involved in mergers during 2006-2022 had a relatively short history, smaller sizes, and greater distances between countries compared to 1966-1990. 
In our estimation results, we observe some transitions of merger incentives. 
All interviewees concurred with these findings. 
However, Sato and Fujita remarked that the complexity of mergers was influenced by multiple industries, including not only the container shipping industry but also shipbuilding and port management in each country. 
Han identified various potential factors driving mergers, such as existing business relationships, the proximity of CEOs or owners, the merger initiator's desire to expand their network, and government support. 
Since no model can capture all potential elements, we believe our model effectively captures data patterns within the container shipping industry, given the available limited data.

\paragraph{Consistency of our counterfactual simulation results}
None of the interviewees expressed surprise at our counterfactual simulation results. 
However, Han noted that the composition of counterfactual mergers could be influenced by specific local policies, such as the level of government intervention. 
In practice, unobserved variables like experience in alliances could also be significant.

\section{Practical Implications, Discussion, and Future Research}\label{sec:practical_implications}

\subsection{Practical Implications}

Our study makes an important contribution to the development of the unified merger list of the container shipping industry and policy discussions for practitioners. 
The data contribution enables practitioners to obtain empirical and historical knowledge on the main container transport markets and to develop a methodology to disentangle merger incentives.
For instance, our analysis corroborates anecdotal evidence indicating that container shipping firms that abstained from joining alliances in the 1980s, and had relatively brief market tenure, primarily engaged in mergers to increase their scale in their home regions. 
Recently, they have tended to merge with companies in distant regions, even when there are no significant differences in size, to expand their container shipping networks.
Additionally, \cite{jeon2022learning} simulates counterfactual and exogenous merger scenarios of specific two firms between 2006 and 2014; however, does not incorporate and explain more than ten merger cases in her model.
Thus, our data contribution complements previous studies institutionally and sheds light on the data patterns and main drivers of mergers in the industry.

Our results also primarily offer managerial insights to chairpersons and logistics managers in the global container shipping industry regarding potential future industry composition. If the current merger trend continues, mergers are likely to involve small and new firms in close countries. Chairpersons should prepare for this potential industrial composition by considering strategies such as building large ships or addressing antitrust concerns. Meanwhile, logistics managers should prepare for profit uncertainty by making informed decisions about shipping routes, considering the involvement of potential merger firms.

Understanding shipping company merger trends is also crucial for shippers, as these mergers often result in route and port call overlaps, necessitating service rationalization. This impact is particularly significant in areas heavily reliant on specific shipping companies, where reduced port call frequency can lead to higher freight rates and increased lead times, posing challenges to maintaining stable supply chains. Shippers may need to reconsider their port choices to ensure supply chain stability.

As a contribution to the policy discussion, our counterfactual finds that the prohibition of mergers between firms in the same country would change the merger configuration significantly. 
This predicts the propagated impact of the prohibition caused by the local competition policies not only on the local markets but also on the global market configuration. 
For example, if Japanese container shipping companies were restricted from merging their container divisions owing to a dominant market share in Japan, they might have pursued mergers with firms in other countries to enhance their shipping network and scale. Such a scenario could have altered the current structure of the container shipping market.

Analyzing historical merger trends provides insights into future patterns. Notably, there is a considerable size disparity between the top five global shipping companies and those below them, indicating ongoing potential for mergers. For example, in regions like South Korea, China, and Taiwan, several small and medium-sized container shipping companies have yet to undergo mergers. The formation of alliances, such as the Korea Shipping Partnership in 2017 and the K Alliance in 2021, highlights the continued potential for mergers in these regions. This ongoing trend underscores the relevance of considering mergers in the shipping industry for industry professionals.

\subsection{Discussion}
We summarize the potential concerns for the data and methodology. 
First, we merge three data sources that record potentially different variables and observations for each regime. 
Thus, we could not check the robustness check on the choice of the regime owing to data limitation and inconsistency, although we believe that our choice of the regime is reasonable for institutional and graphical reasons.
Second, we may face a small sample problem, particularly between 1966 and 1990, although the matching maximum score approach works in a small sample in Monte Carlo simulations \citep{akkus2015ms,otani2021matching_cost}.
Third, we eliminate merger cases that involve firms whose variables are missing in our three data sources. 
This might be because unlike the MDS Transmodal data mentioned above, it does not cover all the routes and vessels in the container trade. This is because the IHS data lacks some historical information on vessel operations, and the HB data collects data on routes and vessels focusing on trunk lines and routes to/from Japan, using operator advertisements. For example, Maersk's merger with Sea-Land and P\&O Containers' merger with Nedlloyd are not in the data and are not reflected in the analysis, as shown in the footnotes in the merger lists in Tables \ref{tb:merger_list_IHS} and \ref{tb:merger_list_HB}.
Fourth, we could not incorporate ship-level, firm-level, or port-level shipping technology,  managers, shareholders, regional industry development, and so on because we could not find these unified data sources. 
Even if all of these variables could be observed, tractability of \cite{fox2018qe}'s approach would be lost.
To the best of our knowledge, our data and findings are entirely novel, shedding light on the historical trends of the global container shipping industry, despite the historical absence of some potential factors.
Fifth, potential effects derived from non-cooperative game theoretical models, that is, synergy effect, competitive power, some gains filling up their missing service network that the merged company owned could not be investigated in our matching model as in \cite{akkus2015ms}. 
Although \cite{fox2019externality} and \cite{uetake2019entry} incorporated the strategic decisions into the TU and NTU matching models in different ways, the externality issues are beyond the state of the methodological literature.\footnote{As discussed in \cite{agarwal2021market}, the condition of the existence of the equilibrium with externality is an active area of theoretical research. The conditions of the uniqueness of the equilibrium are not investigated. }

\subsection{Future Research}

We should discuss possible extensions and some shortcomings of this study. 
As a methodological issue, first, this study focuses on disentangling endogenous merger incentives while ignoring future competition in the market owing to data limitations. 
Thus, a welfare evaluation of the post-merger market was not investigated.
Combining firms' strategic interactions with estimations of demand and supply sides with the endogenous matching merger model remains a challenging and open research question in the field of industrial organization \citep{agarwal2021market}. 
An exceptional study is \cite{igami2019mergers} which construct a stochastic sequential-move game but need monthly-level merger data and allow only a single merger each month. 
Developing their approach might resolve the relationship between matching and competition.
Second, our matching model does not incorporate unobserved heterogeneity which is identified nonparametrically \citep{fox2018jpe}.
Pursuing this direction will require a different econometric approach such as the simulated method of moments in \cite{fox2018jpe} and multiple market data.

\section{Conclusion}\label{sec:conclusion}
We construct a novel unified list of mergers in the global container shipping industry between 1966 (the beginning of the industry) and 2022. 
Combining the list with proprietary data, we construct a structural matching model to describe the historical transition of the importance of a firm's age, size, and geographical proximity on merger decisions. 
We find different transition patterns of the importance of a firm's age, size, and geographical proximity.
In counterfactual simulations, we find that prohibiting mergers between firms in the same country affects the merger configuration of firms involved in prohibited and permitted mergers.

\newpage

\textbf{Acknowledgement} \\
We benefited from anonymous referees and participants at several seminars and conferences. 
We thank Hiroyuki Sato, Yasuhiro Fujita, and Jong Kil Han for their professional comments and Jeremy Fox for introducing methodology.
This study was supported by JSPS KAKENHI Grant Numbers 22K13501, JST ERATO Grant Number JPMJER2301, JSPS KAKENHI Grant Number 24K22604 Grant-in-Aid for Research Activity Start-up, the Port and Harbour Association of Japan Research Grant and Yamagata Maritime Institute Grant.

\bibliographystyle{ecca}
\bibliography{container_merger_data_bib}

\newpage
\appendix
\section{Details of data construction of merger list (Not for publication)}\label{sec:details_of_data_construction}

We notice some differences between our merger list from the data sources and the history of the container shipping industry, in particular, during 1991-2005. 
The differences come from the fact that the data collected by IHS is incomplete, although the data is the best to capture the industry as far as we can use. For reference, We summarize the differences as follows. 

First, as noted by Hiroyuki Sato, former Vice President of Mitsui O.S.K. Lines, numerous small shipping companies not listed in these tables have succumbed to intense competition and declared bankruptcy. This situation arises partly due to the absence of historical vessel data for merged firms in our dataset, which prevents the identification of such mergers. Due to the similar data limitation, as shown in the footnote of our merger lists, our analysis excludes several notable mergers, including the Royal Nedlloyd and Peninsular and Oriental Containers (P\&O Containers) in 1997, NYK and Showa Kaiun in 1998, Maersk and Sea-Land in 1999, and CMA and CGM in 1999. Additionally, we do not account for Ben Line's merger with EAC in 1991, Maersk's acquisition of Ben Line in 1997, Hanjin's merger with Germany's Senator in 1997, CP Ships' merger with Lykes in 1997, the Australian ANZDL in 1998, and Mexican TMM in 1999.

Second, delays and incompleteness in updating operator name information present a significant challenge. This issue primarily stems from the continued use of merged companies' names by the acquiring firm for years following the merger. Several instances exist where we were unable to pinpoint the exact timing of operator name updates in the HB data. Panel (b) of Table \ref{tb:merger_list_HB} highlights mergers with discrepancies that do not align with institutional backgrounds. For instance, APL continued to appear in the HB data until 2017. We also exclude the merger of COSCO and OOCL in 2018 from our analysis, as the merged entity continued to list OOCL as the registered operator name in the HB data until 2022.

\paragraph{Difference between mergers and acquisitions}
The M\&A deals listed in this paper can be classified as either mergers or acquisitions. The majority of the cases in Table \ref{tb:merger_list_CIY} were mergers, with the exception of one case. The other case also neared to merger as the company was purchasing services from a company that was withdrawing from the container industry.

The cases shown in Table \ref{tb:merger_list_IHS} are mainly from the second half of the 1990s onwards, although the majority of cases are classified as acquisitions. There are instances where the acquired company's brand was deemed valuable and therefore retained (APL and CP SHIPS), as well as cases where the company was retained to oversee other operations, such as ship ownership (SVITZER and DANSK). In some instances, the company in question is subsequently absorbed (as was the case with DELMAS and CP SHIPS). Nine cases out of the 18 cases in Table \ref{tb:merger_list_HB} can be classified as acquisitions.

Mergers and acquisitions differ in whether the target company survives, and are essentially different things in terms of corporate decision-making. However, as in the analysis in this paper, as long as the main focus of the analysis is on the scale of service provision and the transport network itself, it can be seen that there is no significant difference between merger and acquisition in the container transport business. Because even in cases where only the container transport business is acquired from the target company, or the buyer company remains as a shareholder or retains the brand, the independence of the transport network is considered to be low.

For example, the decision to entry an alliance is made in a similar manner, regardless of whether it is a merger or an acquisition. In the case of Royal Nedlloyd Lines and P\&O Containers, which merged in 1997, the participating alliances differed. However, following the merger, they became a single company within the alliance. Prior to NOL's acquisition of APL in 1998, the participating alliances differed, but following the acquisition, they became part of the same alliance.

\end{document}

%% file: figuretable/merger_list_CIY.tex
\begin{tabular}[t]{rllrl}
\toprule
ID & Seller & Buyer & Year & Type\\
\midrule
1 & Moore-McCormack Lines Inc & United States Lines & 1970 & acquisition\\
2 & OCL & P\&O Containers & 1986 & merger\\
3 & Franco-Belgian Services & Maersk & 1986 & merger\\
4 & Y-S Line & NLS & 1988 & merger\\
5 & Japan Line & NLS & 1988 & merger\\
6 & KSC & Hanjin & 1988 & merger\\
7 & Finland Steamship & Finnlines & 1990 & merger\\
8 & Atlanttrafik/Barber Blue Sea & Wilhelmsen Lines A/S & 1990 & merger\\
\bottomrule
\end{tabular}

%% file: figuretable/merger_list_IHS.tex
\begin{tabular}[t]{rllrl}
\toprule
ID & Seller & Buyer & Year & Type\\
\midrule
1 & SVITZER AS & A P MOLLER & 1996 & acquisition\\
2 & APL LTD & NEPTUNE ORIENT LINES LTD (NOL) & 1997 & acquisition\\
3 & PRIMA SHIPMANAGEMENT SDN BHD & HALIM MAZMIN GROUP & 1999 & acquisition\\
4 & FARRELL LINES INC & CSAV & 2000 & acquisition\\
5 & OOST ATLANTIC LIJN BV & ATLANTIC HORIZON GROUP & 2001 & acquisition\\
6 & CYPRUS MARITIME CO LTD & CYPRUS SEA LINES SA & 2002 & acquisition\\
7 & DANSK SUPERMARKED INVEST A/S & A P MOLLER & 2003 & acquisition\\
8 & THE PENINSULAR AND ORIENTAL ST & A P MOLLER & 2004 & merger\\
9 & EUROBULK LTD & EUROSEAS LTD & 2005 & acquisition\\
10 & CP SHIPS LTD & HAPAG-LLOYD AG & 2005 & acquisition\\
11 & DELMAS & CMA CGM HOLDING & 2005 & acquisition\\
12 & HORIZON LINES INC & MATSON NAVIGATION CO INC & 2005 & acquisition\\
13 & ROYAL P\&O NEDLLOYD NV & A P MOLLER & 2005 & merger\\
14 & UNITED THAI SHIPPING CORP LTD & IMC SHIPPING CO PTE LTD & 2005 & acquisition\\
\bottomrule
\end{tabular}

%% file: figuretable/merger_list_HB.tex
\begin{tabular}[t]{rllrl}
\toprule
ID & Seller & Buyer & Year & Type\\
\midrule
1 & Cheng Lie & CMA-CGM & 2006 & acquisition\\
2 & Lloyd Triestino & Evergreen & 2006 & merger\\
3 & Norasia & CSAV & 2006 & acquisition\\
4 & MacAndrews & CMA-CGM & 2007 & acquisition\\
5 & Lufeng & Sinotrans & 2008 & merger\\
6 & NEW ONTO SHIPPING & GOTO Shipping International Ltd & 2010 & merger\\
7 & TSK & NYK & 2010 & merger\\
8 & China Navigation & Swire & 2011 & acquisition\\
9 & CCNI & Maersk & 2015 & acquisition\\
10 & CSAV & Hapag-Lloyd & 2015 & acquisition\\
11 & China Shipping & COSCO & 2016 & merger\\
12 & Shanghai Puhai Shipping & COSCO & 2016 & merger\\
13 & UASC & Hapag-Lloyd & 2017 & acquisition\\
14 & KLINE & Ocean Network Express & 2018 & merger\\
15 & MOL & Ocean Network Express & 2018 & merger\\
16 & NYK & Ocean Network Express & 2018 & merger\\
17 & APL & CMA-CGM & 2017 & acquisition\\
18 & Hamburg Sud & Maersk & 2018 & acquisition\\
\bottomrule
\end{tabular}

%% file: figuretable/merger_list_HB_inconsistent.tex
\begin{tabular}[t]{rllrl}
\toprule
ID & Seller & Buyer & Year & Note\\
\midrule
1 & Safmarine & Maersk & 2008 & Merger occurred in 1999\\
2 & Delmas & CMA-CGM & 2016 & Merger occurred in 2005\\
3 & ANL & CMA-CGM & 2022 & Merger occurred in 1998\\
\bottomrule
\end{tabular}

%% file: figuretable/summary_statistics_of_firms_CIY.tex
\begin{tabular}[t]{lrrrrr}
\toprule
  & N & mean & sd & min & max\\
\midrule
Age (Normalized) & 16 & 0.84 & 0.24 & 0.23 & 1.00\\
Size TEU (Normalized) & 16 & 0.23 & 0.29 & 0.01 & 1.00\\
\bottomrule
\end{tabular}

%% file: figuretable/summary_statistics_of_firms_IHS.tex
\begin{tabular}[t]{lrrrrr}
\toprule
  & N & mean & sd & min & max\\
\midrule
Age (Normalized) & 28 & 0.62 & 0.38 & 0.00 & 1.00\\
Size TEU (Normalized) & 28 & 0.15 & 0.26 & 0.00 & 1.00\\
\bottomrule
\end{tabular}

%% file: figuretable/summary_statistics_of_firms_HB.tex
\begin{tabular}[t]{lrrrrr}
\toprule
  & N & mean & sd & min & max\\
\midrule
Age (Normalized) & 42 & 0.54 & 0.30 & 0.11 & 1.00\\
Size TEU (Normalized) & 42 & 0.09 & 0.18 & 0.00 & 1.00\\
\bottomrule
\end{tabular}

%% file: figuretable/maximum_score_estimate.tex
\begin{tabular}[t]{lcccc}
\toprule
Regime &  & 1966-1990 & 1991-2005 & 2006-2022\\
\midrule
 &  &  &  \vphantom{1} & \\
Firm age: $\beta_1$ &  & 1 & 1 & 1\\
Firm size (TEU): $\beta_2$ &  & {}[-9.503,9.475] & {}[9.858,9.858] & {}[-2.013,-2.013]\\
Distance: $\beta_3$ &  & {}[-9.977,-0.687] & {}[-1.550,-1.550] & {}[-0.002,-0.002]\\
 &  &  &  & \\
\% of correct matches &  & 1.000 & 0.923 & 0.981\\
\bottomrule
\end{tabular}

%% file: figuretable/number_of_mergers_counterfactual_upper_bound.tex
\begin{tabular}[t]{lccc}
\toprule
Regime &  & 1991-2005 & 2006-2022\\
\midrule
Matching Num (data) &  & 14 & 21\\
Prop total match (counterfactual/data) &  & {}[1.000,1.000] & {}[1.000,1.000]\\
Prop same match (counterfactual/data) &  & {}[0.214,0.643] & {}[0.095,0.476]\\
\bottomrule
\end{tabular}